\documentclass[prd,preprint,superscriptaddress,preprintnumbers,eqsecnum,showpacs,nofootinbib,nobibnotes]{revtex4}
\usepackage{amsfonts,amsmath,amssymb,bm,natbib}
\usepackage{graphicx} 
\usepackage{mathtools}

\usepackage[utf8]{inputenc}
\usepackage{slashed}

\newcommand{\be}{\begin{equation}}
\newcommand{\bea}{\begin{eqnarray}}
\newcommand{\ba}{\begin{align}}
\newcommand{\ee}{\end{equation}}
\newcommand{\eea}{\end{eqnarray}}
\newcommand{\ea}{\end{align}}
\def\1eq#1{Eq.~(\ref{#1})}

\def\2eqs#1#2{Eqs.~(\ref{#1}) and~(\ref{#2})}
\def\3eqs#1#2#3{Eqs.~(\ref{#1}),~(\ref{#2}) and~(\ref{#3})}

\def\G{\Gamma}

\def\SMs{\mathfrak{s}}

\newcommand\SmallMatrix[1]{{%
\tiny\arraycolsep=0.3\arraycolsep\ensuremath{\begin{pmatrix}#1\end{pmatrix}}}}

\begin{document}

\title{Higgs Potential from Derivative Interactions}
\date{May 8, 2017}

\author{A. Quadri}
\email{andrea.quadri@mi.infn.it}
\affiliation{Dip.~di Fisica, Universit\`a degli Studi di Milano, via Celoria 16, I-20133 Milano, Italy\\
and INFN, Sezione di Milano, via Celoria 16, I-20133 Milano, Italy}

\begin{abstract}
\noindent
A formulation of the linear $\sigma$ model with derivative interactions is studied.
The classical theory is on-shell equivalent to the $\sigma$ model with the standard quartic Higgs potential. 
The mass of the scalar mode only appears in the quadratic part and not in the interaction vertices, unlike in the 
ordinary formulation of the theory. Renormalization of the model is discussed. 
A non power-counting renormalizable extension, obeying the defining functional identities of the theory, is presented.
This extension is physically equivalent to the tree-level inclusion of  a dimension six effective operator $\partial_\mu (\Phi^\dagger \Phi) \partial^\mu (\Phi^\dagger \Phi)$.
The resulting UV divergences are  arranged in  a perturbation series
around the power-counting renormalizable theory. 
%UV completion of the non-power-counting renormalizable model through a symmetric deformation of the 
%propagator of the massive physical scalar is addressed.
The application of the formalism  to the Standard Model in the presence of the dimension-six operator $\partial_\mu (\Phi^\dagger \Phi) \partial^\mu (\Phi^\dagger \Phi)$ is discussed.
\end{abstract}

%Keywords: 

\maketitle

\section{Introduction}

The discovery of the Higgs boson at the LHC~\cite{Chatrchyan:2012xdj,Aad:2012tfa} has by now firmly established the existence
of a scalar particle as a fundamental ingredient of the spontaneous symmetry breaking  (SSB)
mechanism for electroweak theory~\cite{Higgs:1966ev,Higgs:1964pj,Higgs:1964ia,Englert:1964et}.

On the other end, further investigation is required in order  to understand the properties of the SSB potential.
In addition to the Standard Model (SM) quartic Higgs potential, many other possibilities can be considered.
The model-independent approach based on the effective field theory (EFT) technique allows 
to disentangle the phenomenological consequences of higher dimensional operators~\cite{Englert:2014uua,Gripaios:2016xuo}.
The one-loop anomalous dimensions of dimension-six operators have been studied in~\cite{Jenkins:2013zja,Jenkins:2013wua,Alonso:2013hga,Alonso:2014rga,Cheung:2015aba}.
Prospects of measuring anomalous Higgs couplings at the LHC and at future colliders
have been considered e.g. in~\cite{Arkani-Hamed:2015vfh,Contino:2013gna,Contino:2016spe}.

In this paper we discuss the field-theoretical properties of a formulation
of the SSB potential based on higher derivatives interactions~\cite{Quadri:2006hr} that, at  the classical level, is 
physically equivalent to the quartic Higgs potential.

The main advantage of this formulation is that the mass of the physical Higgs excitation 
only enters in the mass term of the physical field and not in the coupling constant (unlike in the 
ordinary quartic  potential).

This has a number of consequences. 
The functional equations governing the theory include the equation of motion for the 
physical massive mode. Let us denote it by $X_2$. 
In addition to the quadratic mass term, one can include in the action a kinetic term for $X_2$ 
$$ \int d^4 x \, \frac{z}{2} \partial_\mu X_2 \partial^\mu X_2 \, .$$
Once such a term is introduced into the classical action, the equation of motion for $X_2$
is modified by a contribution linear in $X_2$ and still survives quantization (due to the fact that 
the breaking is linear in the quantized fields). On the other hand, 
power-counting renormalizability is lost, since
new divergences arise, vanishing at $z=0$.

This remark leads to a perturbative definition of the non power-counting renormalizable theory
as a series expansion around $z=0$, i.e. the coefficients of the expansion in $z$ of the 
amplitudes are fixed in terms of amplitudes of the renormalizable theory at $z=0$.

%A prescription for the evaluation of the UV divergences at $z \neq 0$ can therefore be formulated
%through a rescaling of the $X_2$-lines and an associated mass redefinition in each diagram at $z=0$.
%contributing to a given amplitude. 

%This prescription selects uniquely the  $z$-dependent finite counterterms 
%that one can freely add order by order in the loop expansion, leaving only those allowed in the theory at $z=0$.

By going on-shell and eliminating $X_2$ via the relevant equation of motion 
we obtain the linear $\sigma$ model with a quartic Higgs potential plus 
the dimension six operator $\sim z \partial_\mu \Phi^\dagger \Phi \partial^\mu  \Phi^\dagger \Phi $.

In this model the mass of the physical scalar particle is given by $M^2_{phys} = \frac{M^2}{1+z}$.
The limit where the mass scale of the theory $M$ goes to infinity while keeping $M_{phys}$ fixed 
coincides with the strongly interacting regime $z \rightarrow \infty$.

%In the prescription for the finite parts of the $z$-dependent counterterms outlined before, 
%one gets back the amplitudes of the nonlinear $\sigma$ model in the formulation described in~\cite{Quadri:2006hr}. 

%%% to be moved in another Section
%It should be noticed that, due to the fact that $X_2$ is a SU(2) singlet, the functional symmetries of the theory
%are still preserved if one considers a more general UV-regularised propagator of the $X_2$ field.
%Such a modified propagator is supposed to encode information about a possible UV completion of the theory at a much
%higher scale $\Lambda \gg M_{phys}$. 
%Remarkably enough, one can choose a symmetric (i.e. fulfilling all functional identities of the theory) deformation of the 
%$X_2$-propagator in such a way that power-counting renormalizability is preserved also at $z\neq 0$, at the price
%of introducing at the tree-level a non-local two point 1-PI function for $X_2$.
%The spectrum of the theory is unchanged since there are no further poles in the $X_2$-propagator besides the one 
%at the physical mass $p^2=M^2_{phys}$.
%It should be noticed that power-counting renormalizability can also be recovered by adding further physical
%massive scalars in a way compatible with locality.
%
%These effects
%of would-be new physics at the scale $\Lambda$ can be described via a modification of the two-point sector of the $X_2$-field only,
%while keeping the interactions fixed, unlike in the effective field theory approach, where more and more operators are included, arranged in a power series in $1/\Lambda$.
%%% to be moved in another Section

The $X_2$-equation allows one to control the UV divergences of the one-particle-irreducible (1-PI) Green's functions with $X_2$-legs
in terms of amplitudes with insertions of external sources with a better UV behaviour.

This allows one to disentangle some new relations between 1-PI amplitudes involving  $X_2$-external lines
that are not apparent on the basis of the power-counting of the underlying effective field theory.

One might then conjecture that such relations will translate into consistency conditions between the
Green's functions in the ordinary Higgs EFT once one eliminates $X_2$ through the equations of motion
of the auxiliary fields. This is currently under investigation~\cite{BQ:2017}.

\medskip
The paper is organized as follows. In Sect.~\ref{sec.1} we introduce our notation and discuss
the theory at $z=0$. The BRST symmetry is given and the tree-level on-shell equivalence with the
linear $\sigma$ model in the presence of a quartic Higgs potential is discussed.
In Sect.~\ref{sec:tl} the mechanism 
guaranteeing the on-shell equivalence of the derivative interactions with the usual
quartic Higgs potential is elucidated on some sample tree-level computations.
In Sect.~\ref{sec:ren} the functional identities of the theory are presented and the renormalization of the model
at $z=0$ is carried out. The one-loop divergences are computed  and the on-shell normalization conditions are 
presented.
Sect.~\ref{sec:SM} describes the  Standard Model (SM) action in the derivative representation
of the Higgs potential. The BRST symmetry of the SM is provided.
 In Sect.~\ref{sec:zmod} the non power-counting renormalizable theory at $z\neq 0$ is considered and 
the differential equation defining the Green's functions of the theory as an expansion around $z=0$ is introduced.
In Sect.~\ref{sec:expansion} the UV subtraction of the model at $z\neq 0$ is studied.
The ambiguities in the choice of the finite parts of the counterterms 
 at $z \neq 0$ are related to the insertion
at zero momentum of the quadratic operator $X_2 \square X_2$ in the amplitudes of the 
power-counting renormalizable theory at $z=0$.
In Sect.~\ref{sec:UV} we analyze how non-local symmetric deformations of the $X_2$-propagator
can induce a UV completion of the theory, restoring power-counting renormalizability also at $z\neq 0$, as well as
the UV completion realized through the addition of further physical scalars, while preserving locality of the classical action.
Conclusions are presented in Sect.~\ref{sec:concl}. 
%The Appendices are devoted to some technical aspects: ...

\section{Formulation of the theory}\label{sec.1}

We start from the following action written in components
\begin{align}
S_0 = \int d^4 x \, \Big [ & \frac{1}{2} \partial^\mu \sigma \partial_\mu \sigma + \frac{1}{2} \partial^\mu \phi_a \partial_\mu \phi_a - \frac{M^2}{2} X_2^2 \nonumber \\
& + \frac{1}{v} (X_1 + X_2) \square \Big ( \frac{1}{2} \sigma^2 + v \sigma + \frac{1}{2} \phi_a^2 - v X_2 \Big ) 
 \Big ] \, .
\label{action}
\end{align}
The fields $(\sigma, \phi_a)$ belong to a SU(2) doublet 
\begin{align}
\label{su2.doublet}
\Phi = \frac{1}{\sqrt{2}} 
\SmallMatrix{ i \phi_1 + \phi_2 \\ \sigma + v - i \phi_3} \, ,
\end{align}
$X_2$ is a SU(2) singlet.
$v$ is the scale of the spontaneous symmetry breaking.

The equation of motion of $X_1$ is 
\begin{align}
\frac{\delta S_0}{\delta X_1} = \frac{1}{v} \square \Big ( \frac{1}{2} \sigma^2 + v \sigma + \frac{1}{2} \phi_a^2 - v X_2 \Big ) \, .
\label{x1.eom}
\end{align}
Going on-shell with $X_1$ this yields (neglecting zero modes of the Laplacian\footnote{A rigorous argument of why this is legitimate will be given in Subsection~\ref{sec:off-shell} after Eq.(\ref{eq.a1}).})
\begin{align}
X_2 = \frac{1}{2v } \sigma^2 + \sigma + \frac{1}{2v} \phi_a^2 \, ,
\label{tl.1}
\end{align}
which, substituted into Eq.(\ref{action}), gives 
\begin{align}
\left . S_0 \right |_{\rm on-shell} = \int d^4x \, 
\Big [ \frac{1}{2} \partial^\mu \sigma \partial_\mu \sigma + \frac{1}{2} \partial^\mu \phi_a \partial_\mu \phi_a 
-  \frac{1}{2} \frac{M^2}{v^2} \Big ( \frac{1}{2} \sigma^2 + v \sigma + \frac{1}{2} \phi_a^2 \Big )^2  \Big ] \, ,
\label{tl.2}
\end{align}
i.e. at tree level one finds the ordinary SU(2) linear $\sigma$ model with a quartic Higgs potential of coupling constant $\lambda =  \frac{1}{2} \frac{M^2}{v^2}$. 

It should be remarked that the right sign of the potential, triggering the spontaneous symmetry breaking in Eq.(\ref{tl.2}), 
is dictated by the sign of the mass term of the field $X_2$, which in turn is fixed by the requirement of the absence
of tachyons in the theory.

Notice that by going on-shell with $X_1$, asymptotically $\sigma$ coincides with $X_2$,
as can be seen by taking the linearization of the r.h.s. of Eq.(\ref{tl.1}). In particular,
$\sigma$ acquires a mass $M$, as in Eq.(\ref{tl.2}).

\subsection{Off-shell Formalism}\label{sec:off-shell}

The off-shell implementation of the constraint in Eq.(\ref{tl.1}) 
can be realized {\em \`a la} BRST \cite{Becchi:1974md,Becchi:1975nq,Tyutin:1975qk}.
The construction works as follows~\cite{Quadri:2006hr}. 
One introduces a pair of ghost $c$ and antighost $\bar c$ such that
the BRST variation of the antighost is the constraint:
\begin{align}
& s\bar c =  \Phi^\dagger \Phi - v X_2 - \frac{v^2}{2} = \frac{1}{2} \sigma^2 + v \sigma + \frac{1}{2} \phi_a^2 - v X_2 \, .
\label{brst.I}
\end{align}
The Lagrange multiplier field $X_1$, enforcing the constraint in the action $S_0$, pairs with the ghost $c$ into
a BRST doublet~\cite{Piguet:1995er,Quadri:2002nh,Barnich:2000zw}
\begin{align}
& s X_1 = v c \, , \qquad s c = 0 \, .
\label{brst.II}
\end{align}
A set of variables $u,v$ such that $su = v, sv =0$ is known as a BRST doublet or a trivial pair~\cite{Gomis:1994he} since it does not affect the cohomology $H(s)$ of the BRST differential $s$. 
We recall that
$H(s)$ is the set of local polynomials in the fields and their derivatives such that two polynomials ${\cal I}_1$ and ${\cal I}_2$ are equivalent if and only if they differ by a $s$-exact term: ${\cal I}_1 = {\cal I}_2 + s {\cal K}$ for some ${\cal K}$.
 $H(s)$ identifies the local physical observables of the theory~\cite{Barnich:2000zw,Piguet:1995er}. 
 If one considers local functionals (e.g. the action) and allows for integration by parts, one defines in a similar way
 the cohomology $H(s|d)$ of $s$ modulo the exterior differential $d$~\cite{Barnich:2000zw,Piguet:1995er}.
 $H(s|d)$ controls the local deformations of the classical action (including counterterms) as well as 
 (in the sector with ghost number one) the potential
 anomalies of the theory.

By Eq.(\ref{brst.II}) $X_1$ and $c$ are not physical fields of the theory (as expected, since $X_1$ is a Lagrange multiplier and $c$ is required in the algebraic BRST implementation of the off-shell constraint but should not affect the physics itself).
All other fields are BRST invariant:
\begin{align}
\qquad s \sigma = s \phi_a = s X_2 = 0 \, . 
\label{brst.III}
\end{align}
%
%One recovers BRST invariance under the following BRST transformation
%
%\begin{align}
%\qquad s \sigma = s \phi_a = s X_2 = 0 \, , \\
%& s\bar c =  \frac{1}{2} \sigma^2 + v \sigma + \frac{1}{2} \phi_a^2 - v X_2 \, ,
%\label{brst}
%\end{align}
%
$s$ is nilpotent.

One recovers BRST invariance of the action by adding  to $S_0$ the ghost-dependent term 
\begin{align}
S_{ghost} = - \int d^4x \, \bar c \square c 
\label{ghost.action}
\end{align}
so that the full action of the theory is
\begin{align} 
S = S_0 + S_{ghost} \, .
\label{full.action}
\end{align}
Notice that the ghost is a free field.
It should be stressed that the BRST symmetry $s$ is not associated with a local gauge symmetry of the theory.
It implements algebraically the (SU(2)-invariant) constraint in Eq.(\ref{tl.1}). 

We remark that also the pair $\bar c, {\cal F} = \frac{1}{2} \sigma^2 + v \sigma + \frac{1}{2} \phi_a^2 - v X_2$ forms a BRST doublet according to Eq.(\ref{brst.I}) and therefore drops out of the cohomology $H(s)$. The cohomology $H(s)$, 
respecting all the relevant symmetries of the theory,  is thus given by Lorentz-invariant, global SU(2)-invariant polynomials constructed out of the doublet $\Phi$ and  derivatives thereof, $X_2$ being cohomologically equivalent to $\Phi^\dagger \Phi$ according to Eq.(\ref{brst.I}). This is the cohomology of the linear $\sigma$ model, as expected, since the introduction of the fields $X_1, X_2, \bar c, c$ in order to enforce the constraint in Eq.(\ref{tl.1}) should not alter the physics of the theory.

Since the BRST transformation of the antighost $\bar c$ is non-linear in the quantized fields,
one needs one external source $\bar c^*$ in order to control the renormalization of the BRST variation $s \bar c$.
The latter need to be coupled to the source $\bar c^*$, known as an antifield~\cite{Piguet:1995er,Gomis:1994he}.

Thus the tree-level vertex functional of the theory is finally given by
\begin{align}
\G^{(0)} = \int d^4 x \, \Big [ & \frac{1}{2} \partial^\mu \sigma \partial_\mu \sigma + \frac{1}{2} \partial^\mu \phi_a \partial_\mu \phi_a - \frac{M^2}{2} X_2^2  \nonumber \\
&  - \bar c \square c   + \frac{1}{v} (X_1 + X_2) \square \Big ( \frac{1}{2} \sigma^2 + v \sigma + \frac{1}{2} \phi_a^2 - v X_2 \Big ) \nonumber \\
& + \bar c^* \Big (  \frac{1}{2} \sigma^2 + v \sigma + \frac{1}{2} \phi_a^2 - v X_2 \Big )  \Big ] \, .
\label{vf.1}
\end{align}
Notice that the second line can be rewritten as a $s$-exact term as follows:
\begin{align}
S_{constr} & = \int d^4x \, \Big [  - \bar c \square c   + \frac{1}{v} (X_1 + X_2) \square \Big ( \frac{1}{2} \sigma^2 + v \sigma + \frac{1}{2} \phi_a^2 - v X_2 \Big ) \Big ] 
\nonumber \\
& = \int d^4x \, s ( \frac{1}{v} \bar c \square (X_1 + X_2) ) \, .
\label{vf.constr}
\end{align}
We see that the first line of Eq.(\ref{vf.1}) describes the (SU(2)-invariant) action of the linear $\sigma$-model (in the derivative representation of the potential),
the second (BRST-exact) line the off-shell implementation of the constraint in Eq.(\ref{tl.1}), while
the third line contains the antifield-dependent sector.
 
BRST invariance can be  translated into the following Slavnov-Taylor (ST) identity
\begin{align}
\int d^4 x \, \Big ( v c \frac{\delta \G}{\delta X_1} + \frac{\delta \G}{\delta \bar c^*} \frac{\delta \G}{\delta \bar c} \Big ) = 0
\label{sti}
\end{align}
which holds for the full vertex functional $\G$ (the generator of the 1-PI amplitudes, whose leading order in the loop expansion coincides with $\G^{(0)}$).

The ghost field $c$ has ghost number $+1$, the antighost field $\bar c$ has ghost number $-1$.
All other fields and the external source $\bar c^*$ have ghost number zero.
$\G$ has ghost number zero.

%The BRST differential $s$ is nilpotent. This allows to define the cohomology classes $H(s)$ of
%ocal observables of the model.
%Two operators are physically equivalent if they differ by a $s$-exact term:
%
Some comments are in order.
According to Eq.(\ref{brst.I}) we have
\begin{align}
\Phi^\dagger \Phi  - \frac{v^2}{2} =   v X_2 +  s \bar c \, 
\label{eq.a0}
\end{align}
and thus when $X_2=0$  the operator $\Phi^\dagger \Phi  - \frac{v^2}{2}$ is physically equivalent to the null operator. In this case the theory
reduces to the non-linear $\sigma$ model~\cite{Quadri:2006hr}, enforcing off-shell the non-linear constraint
$\Phi^\dagger \Phi  - \frac{v^2}{2}  = 0$. 

When $X_2$ is different than zero,
the theory has the same degrees of freedom as the linear $\sigma$ model.
Notice in particular that the $X_1$-equation of motion
\begin{align}
\frac{\delta S_0}{\delta X_1} = \frac{1}{v} \square \Big ( \frac{1}{2} \sigma^2 + v \sigma + \frac{1}{2} \phi_a^2 - vX_2 \Big ) = 0
\end{align}
yields the most general solution
\begin{align}
X_2 = \frac{1}{2v} \sigma^2 + \sigma + \frac{1}{2v} \phi_a^2 + \chi
\label{eq.a1}
\end{align}
with $\chi$ a massless degree of freedom satisfying the free Klein-Gordon equation $\square \chi=0$.
Compatibility between Eqs.(\ref{eq.a0}) and (\ref{eq.a1}) entails that $\chi$ must be cohomologically equivalent to the null operator and thus one can safely set $\chi=0$ when going on-shell with the auxiliary field $X_1$.

Since $X_2$ is a scalar singlet, one can add any polynomial in $X_2$ and ordinary derivatives
thereof to $\G^{(0)}$ without breaking BRST symmetry. With the conventions adopted, $\sigma$ 
and $X_2$ have zero vacuum expectation value. This prevents to add the $X_2$-tadpole contribution to the action.
The simplest term is then a mass term for $X_2$ as in  Eq.(\ref{action}).
This turns out to be  compatible with power-counting renormalizability~\cite{Quadri:2006hr}.
 
\subsection{Propagators}

Diagonalization of the quadratic part of $\G^{(0)}$ is achieved by setting
$\sigma = \sigma'+X_1+X_2$.
Then the propagators are
\begin{align}
& \Delta_{\sigma' \sigma'} = \frac{i}{p^2} \, ,  \quad  \Delta_{\phi_a \phi_b} = \frac{i \delta_{ab}}{p^2} \,  ,  \quad \Delta_{\bar c c} = \frac{i}{p^2}  \nonumber \\
& \Delta_{X_1 X_1} = -\frac{i}{p^2} \, , \quad  \Delta_{X_2 X_2} = \frac{i}{p^2 - M^2}  \, .
\label{propagators}
\end{align}
Notice the minus sign in the propagator of $X_1$. This entails that 
the combination $X=X_1+X_2$ has a propagator which falls off as $p^{-4}$ for large momentum:
\begin{align}
\Delta_{XX} = \frac{i M^2}{p^2 ( p^2 - M^2) } \, .
\label{XX.propagator}
\end{align}

\medskip
The physical states of the theory are identified by standard cohomological methods~\cite{Gomis:1994he,Becchi:1996yh}.
The asymptotic BRST charge $Q$ acts on the fields as the linearization of the BRST differential $s$.
$X_1$ is not invariant under $Q$ and thus it does not belong to the physical space ${\cal H} = {\rm Ker }~Q/{\rm Im } ~Q$,
as well as $\sigma'$ (since $[Q, \sigma']=  [Q, \sigma - X_1 - X_2] = - v c$). The combination $\sigma' + X_1$ is BRST invariant,
however it is BRST exact, since it is generated by the variation of the antighost field $\bar c$:
\begin{align}
[ Q, \bar c]_+ = \sigma'+ X_1 
\label{Q.asympt}
\end{align}
and thus it does not belong to ${\cal H}$. The only physical modes are the scalar $X_2$ and the fields $\phi_a$, 
namely the degrees of freedom of  the linear $\sigma$ model.

\section{Tree-level}\label{sec:tl}

\subsection{On-shell amplitudes}

We check in this Section the on-shell equivalence between the theory and the 
linear $\sigma$ model on the tree-level 3- and 4-point amplitudes.

 \subsubsection{3-point amplitude}
 
 The off-shell 3-point amplitude is 
 \begin{align}
 {\cal A}_{X_2 X_2 X_2} = -\frac{i}{v } \sum_{i=1}^3 p_i^2
 \label{3pt.tree.level}
 \end{align}
 where the sum is over the momenta of the particles. By going on shell $p_i^2=M^2$
we get
 \begin{align}
 \left . {\cal A}_{X_2 X_2 X_2 } \right |_{{\rm on-shell}} = -\frac{3 i}{v } M^2 \,, 
 \label{3pt.tree.level.on-shell}
 \end{align}
 which coincides with the amplitude of the SU(2) linear sigma model for the coupling constant $\lambda = \frac{1}{2} \frac{M^2}{v^2}$.
  
 \subsubsection{4-point amplitude}
 
The situation is more involved here. Diagrams contributing with an exchange of a $\sigma'$ and $X_1$ of momentum $q$ sum up to cancel out the unphysical pole at $q^2 = 0$, yielding a finite contribution
\begin{align}
 \left . {\cal A}^{(1)}_{X_2 X_2 X_2 X_2} \right |_{{\rm on-shell}} =
 i \frac{16 M^2}{v^2} \, .
\label{4pt.unphys.on-shell}
\end{align}
Diagrams where a $X_2$ particle is exchanged give in turn
\begin{align}
 \left . {\cal A}^{(2)}_{X_2 X_2 X_2 X_2} \right |_{{\rm on-shell}} =
 - i \frac{19 M^2}{v^2}
 - i \frac{9 M^4}{v^2} \Big ( \frac{1}{s-M^2} + \frac{1}{t - M^2} + \frac{1}{u - M^2} \Big ) \, ,
 \label{4pt.unphys.on-shell.X2}
\end{align}
as a function of the usual Mandelstam variables.

The sum is
\begin{align}
\left . {\cal A}_{X_2 X_2 X_2 X_2} \right |_{{\rm on-shell}} 
=  - i \frac{3 M^2}{v^2}
 - i \frac{9 M^4}{v^2} \Big ( \frac{1}{s-M^2} + \frac{1}{t - M^2} + \frac{1}{u - M^2} \Big ) \, .
\label{4pt.on-shell}
\end{align}
The first term is the one arising in the linear sigma model from the contact four-point vertex,
the last three are those generated by diagrams with the exchange of a propagator and two trilinear couplings.
Notice that the contact interaction contribution is controlled by the sum of two terms, originating both from
${\cal A}^{(1)}$ and ${\cal A}^{(2)}$.

 \section{Renormalization}\label{sec:ren}
 
 \subsection{Functional identities}
 
 In addition to the ST identity in Eq.(\ref{sti}), 
 the theory obeys a set of functional identities constraining the 1-PI Green's functions and their UV divergences:
 \begin{itemize}
 \item the ghost and the antighost equations
 \begin{align}
 & \frac{\delta \G}{\delta \bar c} = -\square c \, , & \frac{\delta \G}{\delta c} = \square \bar c \, .
 \label{gh.ag.eq}
 \end{align}
 These equations imply that the ghost and the antighost fields are free to all order in the loop expansion.
 \item since the ghost is free, one can take a derivative w.r.t. $c$ of the ST identity and use 
 the first of Eqs.(\ref{gh.ag.eq}) to get the $X_1$ equation for the full vertex functional
 \begin{align}
 \frac{\delta \G}{\delta X_1} = \frac{1}{v} \square \frac{\delta \G}{\delta \bar c^*} \, .
 \label{X1.eq.full}
 \end{align}
 %
 %It should be noticed that this is a non-trivial property of the theory, since it relates the renormalization
 %of the operator $\square \Big ( \frac{1}{2} \sigma^2 + v \sigma + \frac{1}{2} \phi_a^2 - vX_2 \Big )$ to the one of
 %the constraint $\frac{1}{2} \sigma^2 + v \sigma + \frac{1}{2} \phi_a^2 - vX_2$.
 %
 \item the shift symmetry
 
The shift symmetry
\begin{align}
\delta X_1(x) = \alpha(x) \, , \quad \delta X_2(x) = - \alpha(x) \, , 
\label{shift.1}
\end{align}
gives
\begin{align}
\frac{\delta \G}{\delta X_1} - \frac{\delta \G}{\delta X_2} = 
 - \square (X_1 + X_2) + M^2 X_2  + v \bar c^* \, .
\label{shift.3}
\end{align} 
The r.h.s. is linear in the quantized fields and therefore the classical symmetry can be extended at the full quantum level.
By using the $X_1$ equation (\ref{X1.eq.full}) into Eq.(\ref{shift.3}) we obtain the $X_2$-equation
\begin{align}
\frac{\delta \G}{\delta X_2} = 
\frac{1}{v} \square \frac{\delta \G}{\delta \bar c^*} 
 + \square (X_1 + X_2) - M^2 X_2  - v \bar c^* \, .
\label{X2.equation}
\end{align}

 \item global SU(2) invariance
 \begin{align}
 \int d^4x \, \Big [ -\frac{1}{2} \alpha_a \phi_a \frac{\delta \G}{\delta \sigma} + 
 \Big ( \frac{1}{2} (\sigma + v) \alpha_a + \frac{1}{2} \epsilon_{abc} \phi_b \alpha_c \Big )
 \frac{\delta \G}{\delta \phi_a} \Big ] = 0 \, .
 \label{global.su2}
 \end{align}
 In the above equation $\alpha_a$ are constant parameters and $\G$ denotes the full 1-PI vertex functional
 (the generator of 1-PI amplitudes). 

 \end{itemize}
 \subsection{Power-counting}
 
 The potentially dangerous interaction terms are the ones involving two derivatives arising from the
 fluctuation around the SU(2) constraint, namely
 \begin{align}
 \int d^4 x \, \frac{1}{v} (X_1 + X_2) \square \Big ( \frac{1}{2} \sigma^2 + \frac{1}{2} \phi_a^2 \Big ) \, .
 \label{r.1}
 \end{align}
 1-PI Green's functions involving external $X_1$ and $X_2$ legs are not independent, 
 since they can be obtained through the functional identities Eqs.~(\ref{X1.eq.full}) and (\ref{X2.equation}) 
 in terms of amplitudes only involving insertions of $\sigma', \phi_a$ and $\bar c^*$.
 For these amplitudes the dangerous interaction vertices in Eq.(\ref{r.1}) are always connected inside loops
 to the combination $X$. Since the propagator $\Delta_{XX}$ falls off as $1/p^4$ for large momenta,
 it turns out that the theory is still renormalizable by power counting.
 
%A formal proof is given in Appendix~\ref{app:pc.ren}.
The UV indices of the fields and the external source $\bar c^*$ are as follows:
$\sigma'$ and $\phi_a$ have UV dimension $1$, $\bar c^*$ has UV dimension $2$.

\subsection{Structure of the counterterms}

We consider the action-like sector independent of $X_1$ and $X_2$, since amplitudes involving
these latter fields are controlled by Eqs.(\ref{X1.eq.full}) and (\ref{X2.equation}). 

Eq.(\ref{global.su2}) entails that the dependence on $\sigma$ and $\phi_a$ can only happen through
action-like functionals invariant under global SU(2) symmetry, namely
\begin{align}
{\cal L}_{ct,1}  = & -{\cal Z} \Big ( \frac{1}{2} \partial^\mu \sigma \partial_\mu \sigma + \frac{1}{2} \partial^\mu \phi_a \partial_\mu \phi_a \Big ) - {\cal M} \Big ( \frac{1}{2} \sigma^2 +  v \sigma + \frac{1}{2} \phi_a^2 \Big ) \nonumber \\ 
& - {\cal G} \Big ( \frac{1}{2} \sigma^2 +  v \sigma + \frac{1}{2} \phi_a^2 \Big )^2 
 - {\cal R}_1 \bar c^*  \Big ( \frac{1}{2} \sigma^2 +  v \sigma + \frac{1}{2} \phi_a^2 \Big ) \, .
\label{ct.1}
\end{align}
There also two invariants depending only on $\bar c^*$, i.e.
\begin{align}
{\cal L}_{ct,2} = -{\cal R}_2 \bar c^* - \frac{1}{2} {\cal R}_3 (\bar c^*)^2 \, .
\label{ct.2}
\end{align}
The most general counterterm Lagrangian at $X_1=X_2=0$ 
is thus given by 
\begin{align}
{\cal L}_{ct} = {\cal L}_{ct,1}+{\cal L}_{ct,2} \, .
\label{ct.full}
\end{align}

Notice the appearance of a quartic potential term absent in the classical action (\ref{action}).
It can be introduced from the beginning into the classical action without violating power-counting renormalizability,
as was done in~\cite{Quadri:2006hr}. 
Notice that if one adds the invariant $\int d^4 x \, {\cal G}^{(0)} \Big ( \frac{1}{2} \sigma^2 + 2 v \sigma + \frac{1}{2} \phi_a^2 \Big )^2 $ at tree level, the physical content of the theory does not change (${\cal G}^{(0)}$ is not an additional physical parameter).
Indeed this term can be rewritten as
\begin{align}
\int d^4 x\, {\cal G}^{(0)} \Big ( \frac{1}{2} \sigma^2 + 2 v \sigma + \frac{1}{2} \phi_a^2 \Big )^2 =
\int d^4 x\, s \Big [  {\cal G}^{(0)}  \bar c \Big ( \frac{1}{2} \sigma^2 + 2 v \sigma + \frac{1}{2} \phi_a^2 + v X_2 \Big ) \Big ] + 
{\cal G}^{(0)}  v^2 X_2^2 \, 
\label{tl.potential}
\end{align}
and this amounts to a redefinition of the mass parameter $M^2 \rightarrow M^2-\frac{1}{2} {\cal G}^{(0)}  v^2$ plus 
a $s$-exact term that does not affect the physics.
The relevant deformations of the functional identities controlling the theory when ${\cal G}^{(0)}$ is non-zero has been given in~\cite{Quadri:2006hr}.

\subsection{One-loop divergences}

The one-loop divergences are controlled by the six coefficients ${\cal Z}^{(1)},
{\cal M}^{(1)}, {\cal G}^{(1)}, {\cal R}^{(j)}$.
Amplitudes are dimensionally regularized in $D$ dimensions.

The amplitude\footnote{We denote by a subscript the fields and external sources w.r.t. which one differentiates,
e.g. $\G_{\bar c^*} = \frac{\delta \G}{\delta \bar c^*}$. It is understood that we set all fields and external sources
to zero after differentiation.} $\G^{(1)}_{\bar c^*}$ fixes ${\cal R}^{(2)} = -\frac{1}{16 \pi^2} \frac{M^2}{4-D}$.
$\G^{(1)}_{\bar c^* \bar c^*}$ fixes ${\cal R}^{(3)} = \frac{1}{4 \pi^2} \frac{1}{4-D}$.
$\G^{(1)}_{\bar c^* \sigma'}$ fixes ${\cal R}^{(1)} = -\frac{1}{8 \pi^2} \frac{M^2}{v^2} \frac{1}{4-D}$.

Moreover
\begin{align}
\left . \G^{(1)}_{\phi_a \phi_b} \right |_{UV div} = \frac{1}{8 \pi^2} \frac{M^4}{v^2} \frac{\delta_{ab}}{4-D} \, .
\label{amp.1}
\end{align}
The divergent part has no momentum dependence. This implies that ${\cal Z}^{(1)}=0$
and ${\cal M}^{(1)} =  \frac{1}{8 \pi^2} \frac{M^4}{v^2} \frac{1}{4-D}$.
Finally from the amplitude $\G^{(1)}_{\sigma'\sigma'}$ one obtains the 
coefficient ${\cal G}^{(1)} = \frac{1}{8 \pi^2} \frac{M^4}{v^4} \frac{1}{4-D}$.

\medskip
Let us now compute the divergences of the one- and two-point functions of $X_1$ and $X_2$. This requires to use 
Eqs.(\ref{X1.eq.full}) and (\ref{X2.equation}).
One finds (we denote the external momenta as arguments of the fields)
\begin{align}
\G^{(n)}_{X_1(0)} = \G^{(n)}_{X_2(0)} = 0 \, , \qquad n \geq 1 \, .
\label{X1-X2.one.point}
\end{align}
Notice that Eqs.(\ref{X1.eq.full})  and (\ref{X2.equation}) holds in the $\sigma-X_1-X_2$ (canonical) basis.
If one performs explicit computations in the most convenient diagonal $\sigma'-X_1-X_2$ basis one needs to take
into account the contributions arising from the field redefinition from $\sigma'$ to $\sigma$, namely on the example
of the one-point functions (we denote by an underline the Green's functions in the diagonal basis whenever they differ from those in the canonical basis)
\begin{align}
\int d^4x \Big ( \G^{(1)}_{\sigma'} \sigma' +  \underline{\G}^{(1)}_{X_1} X_1 + \underline{\G}^{(1)}_{X_2} X_2 \Big ) =
\int d^4x \Big [ \G^{(1)}_{\sigma'} \sigma +  ( \underline{\G}^{(1)}_{X_1} -  \G^{(1)}_{\sigma'} ) X_1 + 
( \underline{\G}^{(1)}_{X_2} -  \G^{(1)}_{\sigma'} ) X_2 \Big ] \, 
\label{one-point.change.var}
\end{align}
so that one finds by explicit computations
\begin{align}
\G^{(1)}_{\sigma'(0)} = \underline{\G}^{(1)}_{X_1(0)} = \underline{\G}^{(1)}_{X_2(0)} =  \frac{1}{16 \pi^2} \frac{M^2}{v} A_0(M^2)  
\label{one-point-diag}
\end{align}
in terms of the standard Passarino-Veltman scalar function $A_0(M^2)$ (we use the  conventions of~\cite{Denner:1991kt,Hahn:2000kx}).
Hence from Eq.(\ref{one-point.change.var})
\begin{align}
& \G^{(1)}_{X_1(0)}  =  \underline{\G}^{(1)}_{X_1(0)} -  \G^{(1)}_{\sigma'(0)} = 0 \, , 
& \G^{(1)}_{X_2(0)}  =  \underline{\G}^{(1)}_{X_2(0)} -  \G^{(1)}_{\sigma'(0)} = 0 \, ,
\label{one-point-canonical.basis}
\end{align}
consistent with Eq.(\ref{X1-X2.one.point}).

\medskip
For the two-point functions in the canonical basis we get
\begin{align}
\G^{(n)}_{X_2(-p) X_2(p)} = \G^{(n)}_{X_1(-p) X_1(p)} = \G^{(n)}_{X_1(-p) X_2(p)} = \frac{1}{v^2} p^4 \G^{(n)}_{\bar c^*(-p) \bar c^*(p)} \, , \qquad n \geq 1 \, ,
\label{2point.X1-X2.functions} 
\end{align}
so that the common UV divergent part is 
$$ \left . \frac{1}{v^2} p^4 \G^{(n)}_{\bar c^*(-p) \bar c^*(p)} \right |_{UV div} = \frac{1}{4 \pi^2} \frac{p^4}{v^2} \frac{1}{4-D} \, . $$

\subsection{Normalization conditions}

The two-point functions of $X_1$ and $X_2$ in the diagonal basis read
\begin{align}
& \underline{\G}^{(n)}_{X_1(-p) X_1(p)} =  \underline{\G}^{(n)}_{X_2 (-p)X_2(p)} = \underline{\G}^{(n)}_{X_1(-p) X_2(p)}  = \frac{p^4}{v^4} \G^{(n)}_{\bar c^*(-p) \bar c^*(p)} - \G^{(n)}_{\sigma'(-p) \sigma'(p)} \, .
\label{2pt-diag.X1X2}
\end{align}
The finite part of the coefficient ${\cal R}^{(n)}_3$ is chosen order by order in the loop expansion
so that
\begin{align}
\left . \underline{\G}^{(n)}_{X_1 (-p)X_1(p)} \right |_{p^2=M^2} = 
\left . \underline{\G}^{(n)}_{X_1(-p) X_2(p)} \right |_{p^2=M^2} = 0 \, .
\label{2pt-on-shell}
\end{align}
This ensures that there is no mixing between $X_1$ and $X_2$ at the pole $p^2 = M^2$.
By Eq.(\ref{2pt-diag.X1X2}) this also implies that
\begin{align}
\left . \underline{\G}^{(n)}_{X_2 (-p)X_2(p)} \right |_{p^2=M^2} =  0 \, ,
\label{2pt-on-shell.X2}
\end{align}
i.e. one is performing an on-shell renormalization, fixing the position of the physical pole of the $X_2$ mode at its tree level value $M^2$.  

The finite part of the coefficient ${\cal R}^{(n)}_2$ is chosen order by order in the loop expansion
in such a way to ensure the absence of a tadpole contribution to $X_2$:
\begin{align}
\left . \underline{\G}^{(n)}_{X_2 (0)} \right |_{p^2=M^2} =   
- \frac{p^2}{v} \left . \G^{(n)}_{\bar c^*(0)} \right |_{p^2=M^2} +  \left . \G^{(n)}_{\sigma'(0)}  \right |_{p^2=M^2} = 0 \, .
\label{X2.tadpole}
\end{align}

The finite part of the coefficient ${\cal R}^{(n)}_1$  is adjusted so that the mixing between $\sigma'$ and $X_2$
vanishes on the pole $p^2= M^2$:
\begin{align}
\left . \underline{\G}^{(n)}_{X_2 (-p) \sigma'(p)} \right |_{p^2=M^2} =   
- \frac{p^2}{v} \left . \G^{(n)}_{\sigma' (-p) \bar c^*(p)} \right |_{p^2=M^2} +  \left . \G^{(n)}_{\sigma'(-p) \sigma'(p)}  \right |_{p^2=M^2} = 0 \, .
\label{X2.sigmaP.mix}
\end{align}
Eqs.(\ref{2pt-on-shell}), (\ref{X2.tadpole}) and (\ref{X2.sigmaP.mix}) ensure that the massive physical scalar mode is described by the field $X_2$ to all orders in the loop expansion.

\medskip
The finite parts of the remaining coefficients ${\cal M}^{(n)}$, ${\cal G}^{(n)}$ and ${\cal Z}^{(n)}$ are chosen in order to ensure the absence of a tadpole for $\sigma'$ and the on-shell normalization conditions for $\sigma'$, namely
\begin{align}
\left . \G^{(n)}_{\sigma'(0)} \right |_{p^2 = 0}= 0 \, , \qquad
\left . \G^{(n)}_{\sigma'(-p)\sigma'(p)} \right   |_{p^2 = 0} = 0 \, , \qquad
\left . \frac{d}{dp^2} \G^{(n)}_{\sigma'(-p)\sigma'(p)} \right   |_{p^2 = 0} = 1 \, .
\label{on-shell.sigmaPrime}  
\end{align}

\section{Inclusion in the Standard Model}\label{sec:SM}

We now discuss how the Higgs potential in the derivative representation is included
into the Standard Model (SM). 

$X_2, X_1$ and the ghosts $c, \bar c$ are invariant under the electroweak gauge group $SU_L(2) \times U_Y(1)$ of weak isospin and hypercharge. The $SU_L(2)$ doublet $\Phi$ transforms as usual under an infinitesimal gauge transformation 
\begin{align}
\delta \Phi =  \Big ( - \frac{i}{2} \alpha_Y + i \frac{\sigma_a}{2} \alpha_a \Big ) \Phi
\end{align}
where $\alpha_a$  and $\alpha_Y$ are the gauge parameters of $SU_L(2)$ and $U_Y(1)$ respectively and 
$\sigma_a$ are the Pauli matrices.
This implies that the couplings with the gauge fields and the fermions are the same as in the SM.

The SM action can be written as the sum of five terms:
\begin{align}
S_{SM} = S_{YM} + S_H + S_F + S_{g.f.} + S_{ghost} \, .
\label{sm.action}
\end{align}
$S_{YM}, S_H, S_F, S_{g.f.}, S_{ghost}$ are respectively the Yang-Mills, the Higgs, the fermion, the gauge-fixing 
and the ghost parts.
$S_{YM}$ and $S_F$ are the same as in the ordinary formulation of the theory  and are given for the sake of completeness in Appendix~\ref{app:A}, where we also collect our notations.

The Higgs part $S_H$ is obtained from Eq.(\ref{action}) upon replacement of ordinary derivatives with the covariant ones
and by adding the Yukawa sector as in the ordinary formulation of the SM:
\begin{align}
S_H = \int d^4 x \, \Big [ &  (D_\mu \Phi)^\dagger D^\mu \Phi  - \frac{M^2}{2} X_2^2 \nonumber \\
& - \sum_{i,j} \Big ( g_{ij} \bar \Psi^L_i \Psi^R_{j,-} \Phi + \tilde g_{ij} \bar \Psi^L_i \Psi^R_{j,+} \Phi^C + h.c. \Big ) \nonumber \\
& + \frac{1}{v} (X_1 + X_2) \square \Big ( \Phi^\dagger \Phi - \frac{v^2}{2} - v X_2 \Big ) 
 \Big ] \, .
 \label{action.higgs}
\end{align}
$\Phi^C$ is the charge conjugated field $\Phi^C = i \sigma^2 \Phi^*$.

Spontaneous symmetry breaking induces a mixing between $\phi_a$ and $\partial A_a$. With the choice 
of the doublet as in Eq.(\ref{su2.doublet}) the mixed bilinear terms read 
\begin{align}
\int d^4x \, \Big ( \frac{g_2 v}{2} \phi_a \partial A_a + \delta_{a3} \frac{g_1 v}{2} \partial B \phi_3 \Big ) \, .
\label{mixing}
\end{align}
$\phi_3$ is coupled to the divergence of the $Z$ field, obtained from the Weinberg rotation as
($A_\mu$ is the photon)
\begin{align} 
A_\mu = c_W B_\mu - s_W A_{3\mu} \, , \qquad Z_\mu = s_W B_\mu + c_W A_{3\mu} \, .
\end{align}
The sine and cosine of the Weinberg angle are given by
\begin{align}
c_W = \frac{g_2}{\sqrt{g_1^2+ g_2^2}}\, , \qquad s_W = \frac{g_1}{\sqrt{g_1^2+ g_2^2}} \, .
\label{wein.angle}
\end{align}
It is also convenient to introduce the charged combinations
\begin{align}
W^\pm = \frac{1}{\sqrt{2}} (A^1_\mu \mp i A^2_\mu) \, , \qquad \phi^\pm = \frac{1}{\sqrt{2}} (\phi^1 \mp i \phi^2) 
\end{align}
The masses of the gauge bosons $W^\pm, Z$ are $M_W = \frac{g_2 v}{2}$, $M_Z=\frac{v}{2}\sqrt{g_1^2+ g_2^2}$.
The mixings in Eq.(\ref{mixing}) are cancelled in a renormalized $\xi$-gauge by choosing
\begin{align}
S_{g.f.} = \int d^4 x \, & \Big ( b^+ {\cal F}^- + b^-{\cal F}^+ + b^Z {\cal F}^Z + b^A {\cal F}^A  + \xi_W b^+ b^- + \frac{\xi_Z}{2} (b^Z)^2 + \frac{\xi_A}{2} (b^A)^2 \Big )
\label{SM.gf}
\end{align}
with the gauge-fixing functions
\begin{align}
{\cal F}^\pm = \partial W^\pm + \xi_W M_W \phi^\pm \, , \quad {\cal F}^Z = \partial Z  + M_Z \xi_Z \phi_3 \, , \quad
{\cal F}^A = \partial A \,.
\end{align}
Finally one constructs the ghost dependent part by summing the SM ghost sector and Eq.(\ref{ghost.action})
\begin{align}
S_{ghost} = \int d^4 x \, \Big ( -{\bar c}^+  \SMs {\cal F}^-  -{\bar c}^- \SMs {\cal F}^+   - {\bar c}^Z \SMs {\cal F}^Z 
- {\bar c}^A \SMs {\cal F}^A  - \bar c \square c \Big ) \, .
\end{align}
In the above equation $\SMs$ is the BRST differential associated with the $SU_L(2) \times U_Y(1)$ electroweak gauge group presented in Appendix~\ref{app.A.BRST}.

The full BRST symmetry of the theory is given by $\tilde s = s + \SMs$. 
$\tilde s^2 = 0$ since both $\SMs$ and $s$ are nilpotent and $s$ and $\SMs$ anticommute,
as a consequence of the fact that the constraint in Eq.(\ref{brst.I}) is invariant under $\SMs$.

\medskip
The physical states of the theory can then be identified as those belonging to the space ${\cal H}_{phys} = {\rm Ker} ~ {\tilde Q}_0/{\rm Im} ~ {\tilde Q}_0$,
where ${\tilde Q}_0$ is the asymptotic charge associated with $\tilde s$:
\begin{align}
& [ {\tilde Q}_0, A_\mu] = \partial_\mu c^A \, , 
[ {\tilde Q}_0, Z_\mu] = \partial_\mu c^Z \, , 
[ {\tilde Q}_0, W^\pm]_\mu = \partial_\mu c^\pm \, , 
\nonumber \\
& [ {\tilde Q}_0, \sigma] = 0 \, , 
[ {\tilde Q}_0, \phi^\pm] = M_W c^\pm \, , 
[ {\tilde Q}_0, \phi_3] = M_Z c^Z \, , \nonumber \\
&  
[{\tilde Q}_0, X_2] = 0 \, ,  [{\tilde Q}_0, X_1] = v c \, , \nonumber \\
& [ {\tilde Q}_0, \Psi^L_i]_+ = 0 \, , [ {\tilde Q}_0, \Psi^R_{i,\sigma}]_+ = 0 \, , \nonumber \\
& [ {\tilde Q}_0, \bar c ]_+ = v (\sigma - X_2) \, ,
[ {\tilde Q}_0, \bar c^A ]_+ = b^A \, ,
[ {\tilde Q}_0, \bar c^Z ]_+ = b^Z \, ,
[ {\tilde Q}_0, \bar c^\pm ]_+ = b^\pm \, , \nonumber \\
& [ {\tilde Q}_0, b^A ] = [ {\tilde Q}_0, b^Z] = [ {\tilde Q}_0, b^\pm ] = 0 \, , \nonumber \\
&  [{\tilde Q}_0, c]_+ =  [{\tilde Q}_0, c^A]_+ =  [{\tilde Q}_0, c^Z]_+ = [ {\tilde Q}_0, c^\pm]_+ = 0 \, .
\end{align}
One sees that the physical states are the physical polarizations of the gauge fields $W^\pm_\mu, A_\mu, Z_\mu$,
the fermion fields and the scalar $X_2$. Notice that $\sigma$ is ${\tilde Q}_0$-invariant, however it belongs to the
same cohomology class of $X_2$, since
\begin{align}
X_2 = \sigma - \frac{1}{v} [ {\tilde Q}_0, \bar c ]_+ \, .
\end{align}
The antighosts and the Nakanishi-Lautrup fields drop out of ${\cal H}_{phys}$ being BRST doublets,
the ghosts and the pseudo-Goldstone bosons do not belong to ${\cal H}_{phys}$, since they also form 
BRST doublets. The standard quartet mechanism~\cite{Becchi:1974xu,Becchi:1975nq,Curci:1976yb,Kugo:1977zq,Ferrari:2004pd} is at work.

\medskip
The equations (\ref{gh.ag.eq}), (\ref{X1.eq.full}) and (\ref{X2.equation})
controlling the dependence of the vertex functional~$\Gamma$ on $X_1, X_2, \bar c$ and $c$
do not change.

We remark that the sector spanned by $X_1,X_2,\bar c$ and $c$ respects custodial symmetry
provided that $X_1,X_2,\bar c$ and $c$ do not transform under the global $SU_L(2)\times SU_R(2)$
group. This can be seen by introducing the matrix
\begin{align}
\Omega = (\Phi^C,\Phi) \, .
\end{align}
Since $\Phi^\dagger \Phi = \frac{1}{2} {\rm Tr} (\Omega^\dagger \Omega)$,
we see that the last line of Eq.(\ref{action.higgs}) is invariant under
the custodial transformation $\Omega '= V_L \Omega V_R^\dagger$, $V_L \in SU_L(2)$, 
$V_R \in SU_R(2)$.

\section{The $z$-model}\label{sec:zmod}

There is a unique term that can be added to the classical action in order to preserve  Eq.(\ref{X2.equation}) 
at the quantum level by deforming its r.h.s. by a linear term in the quantized fields, namely a kinetic term
for $X_2$
\begin{align}
\int d^4 x \, \frac{z}{2} \partial^\mu X_2 \partial_\mu X_2 \, .
\label{kin.term.X2}
\end{align}
By the same arguments leading to Eq.(\ref{tl.2}), upon eliminating $X_2$ by imposing the $X_1$-equation of motion one obtains at tree-level the dimension-six operator
\begin{align}
\int d^4 x \, \frac{z}{v^2} \partial_\mu \Phi^\dagger \Phi \partial^\mu \Phi^\dagger \Phi \, .
\label{dim.six.op}
\end{align}

The functional identities controlling the theory are unchanged with the exception of Eq.(\ref{X2.equation}), which becomes
\begin{align}
\frac{\delta \G}{\delta X_2} = 
\frac{1}{v} \square \frac{\delta \G}{\delta \bar c^*} 
 + \square (X_1 + (1 - z) X_2) - M^2 X_2  - v \bar c^* \, .
\label{X2.equation.z}
\end{align}
Notice that this equation is still valid for the SM with the inclusion of the kinetic term for the $X_2$-field
in Eq.(\ref{kin.term.X2}).
The propagator of $X_2$ is modified as follows
\begin{align}
\Delta_{X_2 X_2} = \frac{i}{(1+z)p^2 - M^2} \, .
\label{X2.prop.x}
\end{align}
Power-counting renormalizability is lost since the cancellation mechanism between the propagator of $X_1$ and $X_2$
is no more at work at $z \neq 0$. Indeed the propagator for the combination $X=X_1 + X_2$ is now
\begin{align}
\Delta_{XX} = \frac{i ( - z p^2 + M^2)}{p^2 [ (1 + z) p^2 - M^2]} \, , 
\label{X.propagator}
\end{align}
so that at $z\neq 0$ the propagator falls off as $p^{-2}$ for large momentum $p$ and therefore cannot compensate 
the contributions from the derivative interaction vertices.

The dependence on $z$ can be controlled by the following differential equation
\begin{align}
\frac{\partial \G}{\partial z} = \int d^4 x \, \frac{\delta \G}{\delta R(x)}
\label{diff.eq.z}
\end{align}
where $R(x)$ is the source coupled in the classical action to the composite operator ${\cal O}(x) = -\frac{1}{2} X_2 \square X_2$.
Notice that the insertion of the operator ${\cal O}(x)$ happens at zero momentum (due to the integration over $d^4x$).

This in turn entails that the Green's functions at $z \neq 0$ can be defined as a perturbative expansion around the power-counting renormalizable model at $z=0$.

\medskip
A comment is in order here. Eq.(\ref{X2.equation.z}) relates the Green's functions with the insertion
of $X_2$-lines with those with the insertion of the source $\bar c^*$, which has a better UV behaviour.
The simplest example is the two-point function for $X_2$, which is obtained by differentiating Eq.(\ref{X2.equation.z})
w.r.t. $X_2$ and then by replacing $\G_{X_2 \bar c^*}$ by using once more Eq.(\ref{X2.equation.z}), this time
after differentiation w.r.t. $\bar c^*$. In momentum space we find
\begin{align}
\G^{(n)}_{X_2(-p) X_2(p)} = \frac{1}{v^2} p^4 \G^{(n)}_{\bar c^*(-p) \bar c^*(p)} \, , \quad n \geq 1 \, .
\label{simpl.ex}
\end{align}
The above equation states that the divergence of the two-point function $\G^{(n)}_{X_2(-p) X_2(p)}$ in 
the canonical basis does not have a constant or a $p^2$-term. Therefore there cannot be any mixing 
between the operator $X_2 \square^2 X_2$ and the operator $X_2^2$ or $X_2 \square X_2$ (off-shell
and by forbidding field redefinitions).
This goes along the same bulk of patterns as those observed in  the one-loop anomalous dimensions 
studied in~\cite{Jenkins:2013zja,Jenkins:2013wua,Alonso:2013hga,Alonso:2014rga,Cheung:2015aba}, although one cannot establish an immediate and straightforward correspondence, due to the fact that 
field redefinitions are used in~\cite{Jenkins:2013zja,Jenkins:2013wua,Alonso:2013hga,Alonso:2014rga,Cheung:2015aba} and moreover those results are valid for one-loop on-shell matrix elements (while 
Eq.(\ref{simpl.ex}) holds off-shell and to all orders in the loop expansion).
We stress the fact that Eq.(\ref{simpl.ex}) is valid at any value fo $z$.

A systematic study of the mixing constraints arising from Eq.(\ref{X2.equation.z}) in the full SM 
with the dimension-six operator induced by the kinetic term in Eq.(\ref{kin.term.X2}) is beyond the scope of the present paper and is currently under investigation~\cite{BQ:2017}.

\section{Expansion around $z=0$}\label{sec:expansion}

Eq.(\ref{diff.eq.z}) states that the contribution of the order $z^n$ of a 1-PI Green's function $\gamma$ 
is obtained by a repeated insertion of the integrated operator $\int d^4x \, {\cal O}(x)$ in the $X_2$-lines of each diagram 
contributing to $\gamma$ in the theory at $z=0$.

\begin{figure}
  \includegraphics[width=50mm]{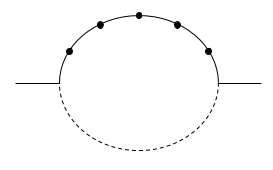}
  \caption{Mass insertions (the dots) on a $X_2$ propagator (drawn as a solid line)}
  \label{fig.1}
\end{figure}

This can be easily understood since the propagator $\Delta_{X_2 X_2}$  can be 
Taylor-expanded according to
\begin{align}
\Delta_{X_2 X_2} & = \frac{i}{(1+z)p^2 - M^2 } = \frac{i}{(p^2 - M^2) \Big [ 1 + \frac{z p^2}{p^2 - M^2} \Big ]} 
= \frac{i}{p^2 - M^2} \sum_{n=0}^\infty (-1)^n \Big ( \frac{z p^2}{p^2 - M^2}  \Big )^n \, .
\label{exp.1}
\end{align}
The term of  order $n$ in the r.h.s. of the above equation is generated by the insertion of 
$n$ vertices $i z p^2$ on the $X_2$-line, namely $n$ insertions of the operators $\int d^4x \, {\cal O}(x)$ at zero external momenta
(see Fig.~\ref{fig.1}).

We remark that the decomposition of the propagator $\Delta_{X_2X_2}$ at $z\neq 0$ can be expressed as the result of repeated mass insertions according to the following identity
\begin{align}
\Delta_{X_2 X_2}  = \sum_{n=0}^\infty (-1)^n z^n \Big [
1 + (1 - \delta_{n,0}) \sum_{k=1}^n %\left ( \begin{matrix} n \cr k \end{matrix} \right )
\dbinom{n}{k}
\frac{M^{2k}}{k} \frac{\partial^k}{\partial (M^2)^k} \Big ] \frac{i}{p^2 - M^2} \, .
\label{exp.2}
\end{align}
\subsection{Two-point function}

The $z$-dependence of the diagrams can be derived according to a simple prescription, namely
one multiplies each diagram in the power-counting renormalizable theory at $z =0$,
with $N$ internal $X_2$-lines and contributing to $\gamma$,  by a prefactor
$(1 + z)^{-N}$ and replaces $M^2 \rightarrow M^2/(1+z)$. 
Then the UV divergences are related directly to those of the 
amplitudes evaluated at $z=0$.

%If one adds finite $z$-dependent counterterms order by order in the loop expansion, one is introducing an additional source of $z$-dependence into the amplitudes that spoils the simple prescription given here. 
%The latter is compatible only with the finite renormalizations of the theory at $z = 0$.

\begin{figure}
  \includegraphics[width=120mm]{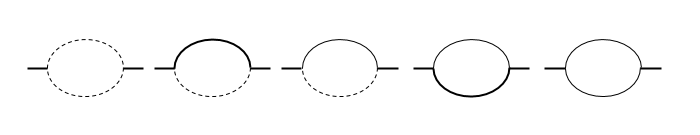}
  \caption{Diagrams contributing to $\G^{(1)}_{\sigma'\sigma'}$  (thin solid lines denote
  the $X_2$  propagator,  thick solid lines the $\sigma'$ propagator. A dashed line denote a $X_1$ propagator)}
  \label{fig.single.diags}
\end{figure}

We consider here as an example the two-point amplitude $\G^{(1)}_{\sigma'\sigma'}$ in the linear $\sigma$ model.
There are five diagrams contributing to this amplitude, depicted in Fig.~\ref{fig.single.diags}.

The first two diagrams are UV convergent, the third and the fourth contain one $X_2$-line, the fifth two
internal $X_2$-lines. This dictates the behaviour of the prefactors $1/(1+z)$ in front of each amplitude.

The physical mass of the canonically normalized $X_2$ field is
\begin{align}
M^2_{phys} = \frac{M^2}{1+z} \, .
\label{a.1}
\end{align}
If one fixes $M_{phys}$ (which, when the model is embedded as the scalar sector
of the electroweak theory, corresponds to the measured Higgs mass), then for amplitudes 
expressed in terms of $M_{phys}$ there are no further sources of $z$-dependence other than the prefactors $1/(1+z)$.

Considering $M$ as the mass of new physics, the limit $M\rightarrow \infty$ is equivalent to $z\rightarrow \infty$.
In this limit the diagrams involving the exchange of internal $X_2$-lines go to zero and one gets back the 
 non-linear $\sigma$ model described in~\cite{Quadri:2006hr}.
 
 %Eq.(\ref{a.1}) gives a very simple formula connecting at tree-level the UV mass scale $M$ with
 %the physical mass $M_{phys}$. For natural values of $z$ around $z=0$ $M \sim M_{phys}$ and the physics is
 %deemed "natural".  

In terms of $M_{phys}$ the one-loop divergence of $\G^{(1)}_{\sigma'\sigma'}$ at finite $z$ is
\begin{align}
\G^{(1) UV}_{\sigma'\sigma'} =  \frac{M_{phys}^2}{8 \pi^2 v^2(4-D) } \Big [ 
3 M_{phys}^2\frac{1-z}{(1+z)^2}
-  p^2 \frac{z}{(1+z)^2} \Big ] \, .
\end{align}
For large $z$ it is of order $1/z$, as expected (since the diagrams 1 and 2 are not  UV divergent and
the remaining three contain at least one $X_2$-line). 
At $z=0$ one recovers the divergence of the power-counting renormalizable theory.

\medskip
%Non power-counting renormalizability at $z\neq 0$ is reflected into the fact that the external source
\subsection{Finite renormalizations}

Consider now in the theory at $z=0$ an amplitude $\gamma$ with superficial degree of divergence $\delta(\gamma)$. 
Let us denote by $\left . \gamma_{R(q_1) \dots R(q_n)} \right |_{q_k=0}$ the diagram with $n$ insertions
of  the operator ${\cal O}$ at zero momentum.
The superficial degree of divergence $\delta(\left . \gamma_{R(q_1) \dots R(q_n)} \right |_{q_k=0})=\delta(\gamma)$, since each of the terms in the series of the r.h.s. of  Eq.(\ref{exp.1}) tends to $1$ for large $p$ and thus does 
not alter the overall behaviour as $1/p^2$ of the propagator $\Delta_{X_2 X_2}$.

This implies that in the power-counting renormalizable theory at $z=0$
the external source $R$ has UV dimension zero.

Consequently when one allows in the theory at $z=0$ for the insertion of the operator $\int d^4x \, {\cal O}(x)$,  the most general form of the counterterms is no more given by Eqs.(\ref{ct.1}) and (\ref{ct.2}).

The problem of identifying the counterterms in the power-counting renormalizable theory at $z=0$ 
in the presence of the source $R(x)$
amounts to find all possible Lorentz-covariant, global SU(2)-invariant local monomials in the fields, the external sources and their derivatives of dimension $\leq 4$.

We can safely disregard all monomials involving derivatives of $R(x)$, since we are only interested in zero momentum insertions. 
Then Eq.(\ref{ct.full}) is modified as follows:
\begin{align}
{\cal L}_{ct,R} = 
& {\cal A} - {\cal Z}_R \Big ( \frac{1}{2} \partial^\mu \sigma \partial_\mu \sigma + \frac{1}{2} \partial^\mu \phi_a \partial_\mu \phi_a \Big ) - {\cal M}_R \Big ( \frac{1}{2} \sigma^2 +  v \sigma + \frac{1}{2} \phi_a^2 \Big ) \nonumber \\ 
& - {\cal G}_R \Big ( \frac{1}{2} \sigma^2 +  v \sigma + \frac{1}{2} \phi_a^2 \Big )^2 
 - {\cal R}_{1,R} \bar c^*  \Big ( \frac{1}{2} \sigma^2 +  v \sigma + \frac{1}{2} \phi_a^2 \Big ) 
 -{\cal R}_{2,R} \bar c^* - \frac{1}{2} {\cal R}_{3,R} (\bar c^*)^2 \, ,
\end{align}
where
${\cal Z}_R, {\cal M}_R, {\cal G}_R$ and ${\cal R}_{j,R}$, $j=1,2,3$ 
(which we collectively denote by ${\cal K}_{l,R}$, with ${\cal K}_l$ standing for the corresponding
quantities in Eqs.(\ref{ct.1}) and (\ref{ct.2})) become analytic functionals of the 
integrated source $R(x)$:
\begin{align}
{\cal K}_{l,R} = {\cal K}_l+ \sum_{k=1}^\infty \frac{1}{k!} \int \, \prod_{i=1}^k d^4x_i \, K_l^k R(x_1) \dots R(x_k) 
\end{align}
while
$${\cal A} =  \sum_{k=1}^\infty \frac{1}{k!} \int \, \prod_{i=1}^k d^4x_i \, A^k R(x_1) \dots R(x_k) \,$$ 
controls the renormalization of the ${\cal O}$-insertions into vacuum amplitudes.
%
%${\cal A}$ controls the renormalization of the vacuum amplitude (no insertions of fields and the source
%$\bar c^*$) at $z\neq 0$.

The finite parts of the coefficients $A^k,K_l^k$ are unconstrained by the symmetries of the theory and
have to be chosen by an (infinite) set of normalization conditions. 
This is the counterpart at $z=0$ of the ambiguities induced by the loss
of power-counting renormalizability at $z\neq 0$.

\section{UV completion}\label{sec:UV}

Under the assumption that all scalar fields should obey at the classical level 
asymptotically Klein-Gordon equations of motion,
the most general $X_2$-equation is given by Eq.(\ref{X2.equation.z}) and includes the kinetic term controlling the 
violation of power-counting renormalizability.
The fact that $X_2$ is a SU(2) singlet  entails however on symmetry grounds 
that more general quadratic terms in the 
tree-level action can be considered without violating the defining functional identities of the theory. 

These terms might capture some features of the UV completion of the model from a more fundamental theory, valid at 
a much higher scale $\Lambda$.
This possibility is a peculiar feature of the higher derivative formulation where the physical parameters of the theory are embodied in the two-point sector.

As an example, 
the bad UV behaviour of the $X$ propagator at $z \neq 0$ can be regularized  by the following choice of the
$X_2$-propagator
\begin{align}
\Delta_{X_2 X_2} = \frac{i}{(1+z)p^2 - M^2} - \frac{i}{\Lambda^2} \Big [ \exp \Big ( - \frac{z}{z+1} \frac{\Lambda^2}{p^2} \Big ) - 1 \Big ] \, .
\label{mod.X2.prop}
\end{align}
For large momenta $\Delta_{X_2 X_2}$ goes as
\begin{align}
\Delta_{X_2 X_2} \sim_{p \rightarrow \infty} \frac{i}{p^2} + i \frac{M^2 - \frac{\Lambda^2}{2} z^2}{(1+z)^2 p^4}
\end{align}
so that $\Delta_{XX}$ goes as $p^{-4}$, making the derivative interaction terms harmless.
Of course such a deformation is not unique (for instance the replacement $\exp~( - \frac{z}{z+1} \frac{\Lambda^2}{p^2}  )\rightarrow \exp~( - z \frac{\Lambda^2}{(1+z) p^2 - M^2} ) $ into Eq.(\ref{mod.X2.prop}) would also work).

This implies that one can regularize the propagator of the $X_2$ field in the UV in such a way to preserve power-counting renormalizability without destroying the symmetries of the model and without introducing new physical poles in the spectrum.

%%% added on August 27, 2016

Power-counting renormalizability at $z \neq 0$ can also be restored by adding new physical SU(2)-invariant modes.
For that purpose consider the following propagator for the field $X_2$ (now a field with a non-trivial
K\"allen-Lehmann spectral density):
\begin{align}
\Delta_{X_2X_2} =  \frac{i}{(1+z)p^2 - M^2} + \sum_{j=1}^N \frac{i}{z_j p^2 - M_j^2} \, ,
\label{prop.kl.2}
\end{align}
where $z_j >0$. The propagator in Eq.(\ref{prop.kl.2}) describes a set of physical resonances at mass
$M^2/(1+z)$ and $M_j^2/z_j$.
A suitable choice of the $z_j$ allows to cancel the UV behaviour in $1/p^2$ of 
$\Delta_{XX}$, therefore re-establishing power-counting renormalizability.
For simplicity we choose all the $z_j$ to have the same common value $z_c$.
Then the leading order in $1/p^2$ of the propagator $\Delta_{XX}$  is cancelled provided that one chooses
\begin{align}
z_c = N \frac{1+z}{z} = N \frac{M^2}{M^2 - M^2_{phys}} \, .
\end{align}
The positivity condition on $z_c$ yields $M>M_{phys}$.
%%%

Both with a UV-regularized propagator and with the addition of further resonances at higher masses, the $X_2$-equation now takes the form (written for convenience in the Fourier space)
\begin{align}
\frac{\delta \G}{\delta X_2(p)} = 
- \frac{1}{v} p^2 \frac{\delta \G}{\delta \bar c^*(p)} 
 -p^2 X_1 + {\cal K}(p) X_2(p)  - v \bar c^*(p) \, ,
\label{X2.equation.z.uv}
\end{align}
where 
\begin{align}
{\cal K}^{-1}(p) = - i \Delta_{X_2 X_2}(p) \, .
\end{align}
The r.h.s. of Eq.(\ref{X2.equation.z.uv}) is still linear in the quantized fields.

It is suggestive that such pieces of information about the UV completion can be embodied  into the two-point function of the $X_2$ scalar, without the need to modify the interaction sector of the theory.

\medskip
Linearity of Eq.(\ref{X2.equation.z.uv}) imposes strong constraints on the allowed potential.
As an example, let us consider the theory with two physical massive scalars, i.e. $X_2 = \chi_1 + \chi_2$
where $\chi_1$ has mass $M/\sqrt{1+z}$ and $\chi_2$ has mass $M_1/\sqrt{z_1}$.
Since both $\chi_1$ and $\chi_2$ are BRST-invariant, bilinear mixing terms involving these fields can be removed
and the quadratic part of the action reads
\begin{align}
S_2 = \int d^4x \, \Big ( & - \frac{1}{2} \sigma' \square \sigma' + \frac{1}{2} X_1 \square X_1  -\frac{z_1}{2} \chi_1 \square \chi_1 - \frac{M_1^2}{2} \chi_1^2 - \frac{1+z}{2} \chi_2 \square \chi_2 - \frac{M^2}{2} \chi_2^2 
\Big ) \, .
\end{align}
The propagator of $X_2$ is then given by Eq.(\ref{prop.kl.2}). The interaction terms are obtained by replacing $X_2 = \chi_1 + \chi_2$. 
The inclusion of the field $\chi_1$ allows to reproduce the Higgs singlet model (HSM), where in addition to the Higgs doublet the real  scalar singlet $\chi_1$ is introduced \cite{Profumo:2007wc,Noble:2007kk,Ashoorioon:2009nf,Espinosa:2011ax,Chen:2014ask,Kanemura:2016lkz,Kanemura:2015fra,Fuyuto:2014yia}.
The most general tree-level potential compatible with SU(2) symmetry and of dimension $\leq 4$ is
\begin{align}
V = - v \mu_\Phi^2 X_2 + \frac{M_2^2}{2} X_2^2 + v \mu_{\Phi \chi} X_2 \chi_1 + 
v^2 \frac{\lambda_{\Phi \chi}}{2} X_2 \chi_1^2 + \mu_{\chi}^3 \chi_1 + 
\frac{M_1^2}{2} \chi_1^2 + \frac{\mu'_\chi}{3} \chi_1^3 + \frac{\lambda_\chi}{4} \chi_1^4
\label{potential.hsm}
\end{align}
Going on-shell with $X_1$ one again finds $X_2 = \frac{1}{v} \Phi^\dagger \Phi$ and
upon substitution in Eq.(\ref{potential.hsm}) we obtain the standard formulation
of the HSM potential. The coefficient of the quartic term $(\Phi^\dagger \Phi)^2$ is
$\lambda_\Phi = \frac{M_2^2}{2 v^2}$.

Imposing linearity of the $X_2$-equation yields the constraint
\begin{align}
\lambda_{\Phi\chi} = 0
\label{mixed.constr}
\end{align}
Provided that $\lambda_\chi >0$, positivity of $M_2$ and Eq.(\ref{mixed.constr}) 
ensures the fulfillment of the vacuum stability condition, namely
$\lambda_\Phi >0 \, , \lambda_\chi >0,  4 \lambda_\Phi \lambda_\chi > \lambda_{\Phi \chi}^2$~\cite{Fuyuto:2014yia}.

\section{Conclusions}\label{sec:concl}

The linear $\sigma$ model can be formulated in such a way that the mass of the physical scalar 
particle only appears in the quadratic part of the action (and not also in the potential coupling)
by introducing suitable derivative interactions 
and additional unphysical fields pairing into BRST doublets.

In the present paper we have given all technical tools required to study such a model and its 
extension to the SM.

Power-counting renormalizability has been established and the BRST symmetry guaranteeing the cancellation of the unphysical degrees of freedom has been given.

The 1-PI amplitudes involving the physical massive scalar field $X_2$ are determined 
in terms of external sources with a better UV behaviour
by a functional identity implementing
at the quantum level the equation of motion for $X_2$. 

Remarkably, such an identity admits a unique deformation, given by a kinetic term for $X_2$ whose
coefficient we have denoted by $z$. 

Once such a kinetic term is introduced into the classical action, power-counting renormalizability is lost.
Violation of power-counting renormalizability is controlled by the parameter $z$. At $z=0$ we recover the original
power-counting renormalizable model.

The amplitudes of the full theory can be expanded as a power series in $z$ with coefficients given by diagrams of the original theory at $z=0$. Each order in $z$ is associated to an additional zero-momentum insertion of the 
kinetic operator into $X_2$-lines. 

This property can be used to express the structure of the counterterms of the non power-counting renormalizable
theory at $z\neq 0$ in terms of analytic functionals of the integrated source $R(x)$ coupled to 
the kinetic term of the $X_2$ field. 

%The theory at $z\neq 0$ comes with a natural prescription for the evaluation of its amplitudes: the latter are given by diagrams in the $z=0$ theory suitably rescaled by factors of $1+z$ according to the number of $X_2$-lines of each diagrams, modulo a redefinition of the mass of the $X_2$-field. At fixed physical mass of the scalar particle, 
%the dependence on $1+z$ only happens through the overall rescaling of the single diagrams.

%The requirement that such a prescription is not spoiled by finite renormalizations yields a consistent choice of the finite counterterms order by order in the loop expansion to all orders in $z$. 

The theory studied in this paper provides a novel example of a non power-counting renormalizable model,
defined as a series expansion around the power-counting renormalizable theory
at $z=0$.

The proposed representation of the Higgs potential does not spoil any of the SM symmetries as well as the
custodial symmetry, in the limit where the $U_Y(1)$ coupling constant vanishes. 

The $X_2$-functional equation
at $z \neq 0$ holds in the SM deformed by the kinetic term for $X_2$.
This paves the way for a rigorous all-orders algebraic study of the renormalization properties
of the effective operator $\partial_\mu (\Phi^\dagger \Phi) \partial^\mu (\Phi^\dagger \Phi)$
in the Higgs EFT. 

\section*{Acknowledgments}

The Author wishes to thank the Service de Physique Th\'eorique et Math\'ematique at the ULB Brussels, where part of this work was carried out, for the warm hospitality. Useful comments by D.~Binosi are gratefully acknowledged.

\appendix

\section{Electroweak SM Lagrangian}\label{app:A}

The generators of the weak isospin $SU_L(2)$ group are $I_a = \frac{\sigma_a}{2}$ where $\sigma_a$, $a=1,2,3$ denote the Pauli matrices.

The Yang-Mills part $S_{YM}$ of the SM Lagrangian is
\begin{align}
S_{YM} = \int d^4 x \, \Big ( -\frac{1}{4} G_{a\mu\nu} G^{\mu\nu}_a - \frac{1}{4} F_{\mu \nu} F^{\mu\nu} \Big )
\label{SM.YM}
\end{align}
where the field strength $G_{a\mu\nu}$ is given in terms of the non-Abelian gauge fields 
$A_{a\mu}$ by ($f_{abc}$ are the $SU_L(2)$ structure constants)
\begin{align}
G_{a\mu\nu} = \partial_\mu A_{a\nu} - \partial_\nu A_{a\mu} + g_2 f_{abc} A_{b\mu} A_{c\nu}
\end{align}
and the Abelian $U_Y(1)$ field strength $F_{\mu\nu}$ is ($B_\mu$ is the hypercharge $U_Y(1)$ vector field)
\begin{align}
F_{\mu\nu} = \partial_\mu B_\nu - \partial_\nu B_\mu \, .
\end{align}
$g_1,g_2$ are the $U_Y(1)$ and $SU_L(2)$ coupling constants respectively.

The fermionic part $S_F$ is 
\begin{align}
S_F = \int d^4x \, \Big ( i \sum_i   \bar \Psi^L_i \slashed{D} \Psi^L_i + i \sum_{i,\sigma} \bar \Psi^R_{i,\sigma}  \slashed{D} \Psi^R_{i,\sigma} \Big )
\end{align}
where the sum is over all  left fermionic doublets and the right singlets.

The index $\sigma$ runs over the right fermion fields corresponding 
to the two components of the associated left doublet, namely if $\Psi_L = \SmallMatrix{ \nu_{L} \\ e_{L}}$,
then $\Psi_{R,+} = \nu_R \, , \Psi_{R,-}= e_R$.

\medskip
The covariant derivative is defined by
\begin{align}
D_\mu = \partial_\mu - i g_2 I_a A_{a\mu} + i g_1 \frac{Y}{2} B_\mu
\end{align}
where $Y$ is the hypercharge. The electric charge is related to $Y$ by the Gell-Mann-Nishijima formula 
$Q=I_3 + \frac{Y}{2}$.

\subsection{BRST Symmetry}\label{app.A.BRST}

We  collect here the BRST symmetry of the SM. $c_a$ are the $SU_L(2)$ ghosts,
$c_0$ is the $U_Y(1)$ ghost:
\begin{align}
& \SMs A_{a\mu} = \partial_\mu c_a + g_2 f_{abc} A_{b\mu} c_c \, , \nonumber \\
& \SMs B_\mu = \partial_\mu c_0 \, , \nonumber \\
& \SMs \Phi =  \Big ( - i g_1 \frac{1}{2} c_0 + i g_2 \frac{\tau_a}{2} c_a \Big ) \Phi \, , \nonumber \\
& \SMs \Psi^L_i = \Big ( - i g_1 \frac{Y^L_i}{2} c_0 + i g_2\frac{\tau_a}{2} c_a \Big ) \Psi^L_i \, , 
\quad \SMs \Psi^R_{i,\sigma} = - i g_1 \frac{Y^R_{i,\sigma}}{2} c_0 \Psi^R_{i,\sigma} \, , \nonumber \\
& \SMs X_1 = \SMs X_2 = \SMs \bar c = \SMs c = 0 \, , \nonumber \\
& \SMs c_a = -\frac{g_2}{2} f_{abc} c_b c_c \, , \quad \SMs c_0 = 0 \, .
\label{brst.SM}
\end{align}

The ghosts in the physical basis of the fields $W^\pm_\mu, A_\mu, Z_\mu$ are obtained by
\begin{align}
c^\pm = \frac{1}{\sqrt{2}} (c^1 \mp i c^2) \, , \quad c_A = c_W c_0 - s_W c_3 \, , \quad c_Z = s_W c_0 + c_W c_3 \, ,
\end{align}
where $c_W,s_W$ are the cosine and sine of the Weinberg angle (see Eq.(\ref{wein.angle})).
The Nakaniski-Lautrup fields in Eq.(\ref{SM.gf}) form BRST doublets with the antighosts:
\begin{align}
\SMs {\bar c}^\pm = b^\pm \, , \quad \SMs b^\pm = 0 \, , \quad \SMs {\bar c}^A = b^A \, , \quad \SMs b^A = 0 \, , 
\quad \SMs {\bar c}^Z = b^Z \, , \quad \SMs b^Z =0 \, . 
\end{align}

\end{document}